\documentclass[pre,aps,twocolumn]{revtex4}

\usepackage{graphicx}
\usepackage{amsmath}
\usepackage{amssymb}
\usepackage{ifthen}
\usepackage{booktabs}

\usepackage{graphicx}% Include figure files
\usepackage{dcolumn}% Align table columns on decimal point
\usepackage{bm}% bold math
\usepackage{longtable}

\newcommand{\bb}{\begin{equation}}
\newcommand{\ee}{\end{equation}}
\newcommand{\ba}{\begin{eqnarray*}}
\newcommand{\ea}{\end{eqnarray*}}
\newcommand{\rhor}{\rho({\bf r})}
\newcommand{\dd}{{\rm d}}
\newcommand{\rr}{{\mathbf r}}
\newcommand{\dr}{{\rm d}{\bf r}}

\begin{document}

\title{Complete wetting near an edge of a rectangular-shaped substrate}

\author{Alexandr \surname{Malijevsk\'y}}
\affiliation{
 {Department of Physical Chemistry, ICT Prague, 166 28 Praha 6, Czech Republic}
}

\begin{abstract}
We consider fluid adsorption near a rectangular edge of a solid substrate that interacts with the fluid atoms via long range (dispersion) forces. The curved
geometry of the liquid-vapour interface dictates that the local height of the interface above the edge $\ell_E$ must remain finite at any subcritical
temperature, even when a macroscopically thick film is formed far from the edge. Using an interfacial Hamiltonian theory and a more microscopic fundamental
measure density functional theory (DFT), we study the complete wetting near a single edge and show that  $\ell_E(0)-\ell_E(\delta\mu)\sim\delta
\mu^{\beta_E^{co}}$, as the chemical potential departure from the bulk coexistence $\delta\mu=\mu_s(T)-\mu$ tends to zero. The exponent $\beta_E^{co}$ depends on
the range of the molecular forces and in particular $\beta_E^{co}=2/3$ for three-dimensional systems with van der Waals forces. We further show that for a
substrate model that is characterised by a finite linear dimension $L$, the height of the interface deviates from the one at the infinite substrate as
$\delta\ell_E(L)\sim L^{-1}$ in the limit of large $L$. Both predictions are supported by numerical solutions of the DFT.
\end{abstract}

\maketitle

\section{Introduction}

It is well known that the adsorption properties of solid substrates strongly depend on the substrate geometry. In particular, the nature of pertinent surface
phase transitions on non-planar substrates may qualitatively differ from those on planar walls. The surface geometry can have a profound influence on the
location of the phase transitions, their order, and the values of the critical exponents, and it can even induce entirely new interfacial phase transitions and
fluctuation effects \cite{cacamo, bonn, saam, hauge, rejmer, wood, nature, silvestre}. Recent theoretical studies also have revealed new examples of surprising
connections between adsorption in different geometries \cite{cov1, cov2}. These findings are not only interesting in their own rights but also have useful and
far-reaching consequences for applications that require the design of modified surfaces, whose adsorption properties can be sensitively controlled at the
nanoscale. Indeed, recent advances in nano-lithography have opened up an entirely new area of research with exciting implications for modern technologies
\cite{whitesides, quere, rauscher} that address the properties of fluids that are geometrically constrained to a molecular scale. Examples of the products of
this sort of innovation include self-cleaning materials \cite{selfclean}, responsive polymer brushes \cite{brush} or ``lab-on-a-chip'' devices \cite{chips}.

A prerequisite to these applications is a detailed description of fluid adsorption on structures of the most fundamental non-planar geometries. This paper
focuses on the adsorption of a simple fluid near a substrate edge. In the simplest case of a single edge, the substrate geometry can be characterised by an
internal angle $\phi<\pi$, where two semi-infinite planes meet. This (convex) object should be distinguished from a (concave) linear wedge model, because the
fluid behaviours on these two substrates are strikingly different. While the wedge geometry promotes fluid condensation and shifts the temperature where
macroscopic coverage occurs below the wetting temperature $T_w$ of a corresponding planar wall \cite{rejmer, wood}, the presence of the substrate edge implies
that the height of the liquid-vapour interface above the edge $\ell_E$ remains finite at any subcritical temperature, even when the interface far from the edge
$\ell_\pi$ unbinds from the wall. This suppression occurs because of the surface free energy cost, that must be paid for interface bending above the edge,
similarly to adsorption on a spherical wall where the growth of an adsorbed film is restricted by the Laplace pressure arising from the curved liquid-vapour
interface \cite{stewart, nold}.

A proper understanding of how the presence of the edge affects the wetting properties of the wall is important to obtain a comprehensive picture of adsorption on
structured (or sculpted) surfaces. Recently, theoretical and experimental studies have shown that a planar wall etched with an array of rectangular grooves
exhibits more adsorption regimes than the simple flat wall \cite{cov2, bruschi1, bruschi2, tasin, hofmann, checco, mal}. A recent density functional (DFT) study
\cite{mal} revealed that hydrophilic grooved surfaces experience the wetting transition at temperature $T>T_w$, which is in contrast with the predictions based
on macroscopic approaches, such as the Wenzel model, predicting that surface corrugation promotes the surface's wetting properties \cite{quere}. Furthermore, the
regimes that are characterised by the formation of a laterally inhomogeneous film with the interface pinned at the groove edges and followed by a discontinuous
unbending \cite{unbending} of the interface have been observed. In these cases, the presence of the groove edges plays a crucial role and the explanation of
these phenomena is incomplete without our knowledge of what occurs in the immediate vicinity of an isolated edge.

\begin{figure*}
\includegraphics[width=0.45\textwidth]{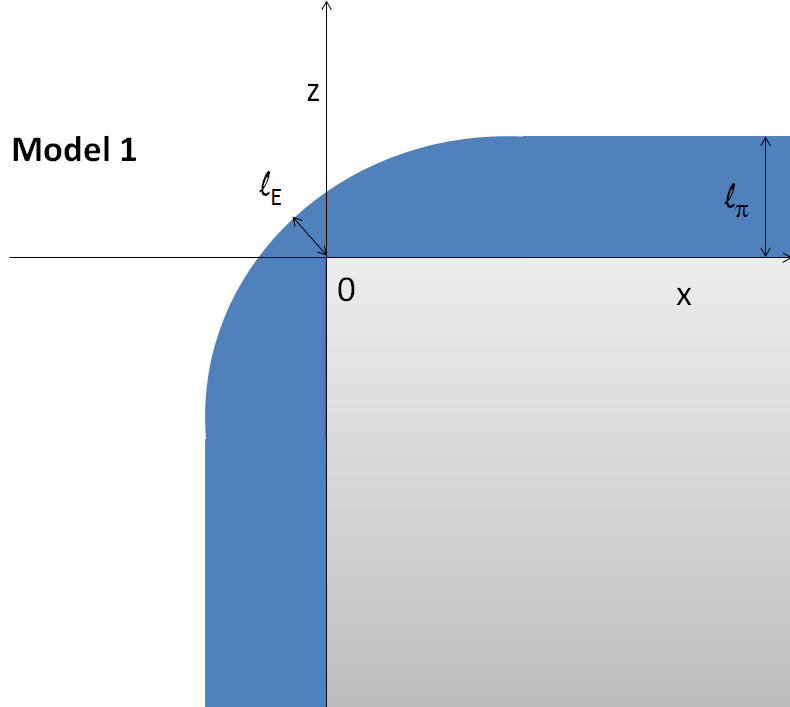} \hspace*{0.5cm}\includegraphics[width=0.45\textwidth]{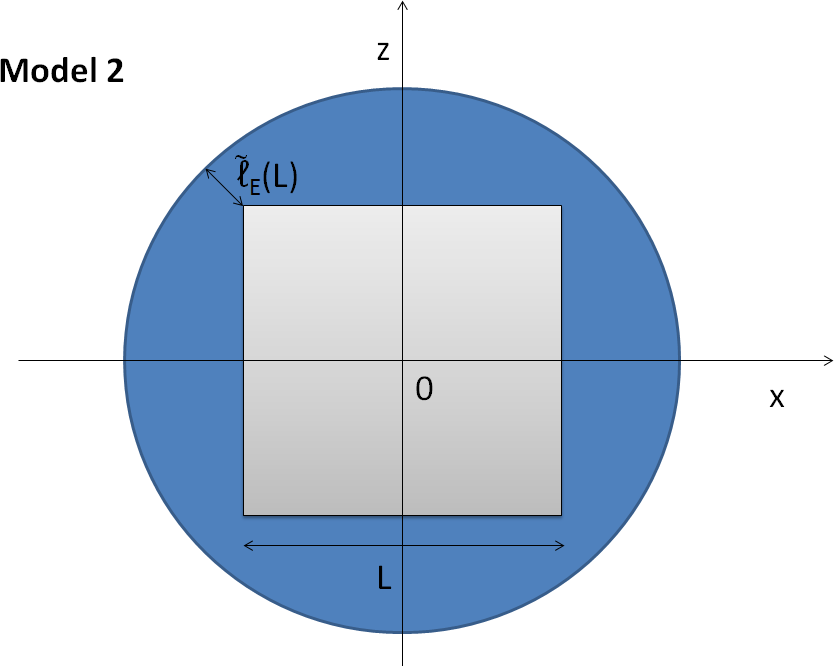}
\caption{Left: A sketch of Model 1 in the $x$-$z$ projection. The liquid film thickness above the edge of the substrate is $\ell_E$, while far away from the edge
the film thickness approaches the value $\ell_\pi$ corresponding to an adsorbed layer above a planar wall. Right: A sketch of Model 2 in the $x$-$z$ projection.
The liquid film thickness above each of the four edges of the substrate is $\ell_E(L)$, which now depends on the linear dimension of the wall $L$. In both cases,
the wall is infinitely long along the Cartesian $y$-coordinate.} \label{sketch}
\end{figure*}

A study by Parry et al. \cite{parry_apex} provides a description of the adsorption near an edge that focuses on the limit of $\phi\to\pi$ and shows a connection
between complete wetting near a shallow edge and critical wetting on a planar wall. Here, motivated by the previously mentioned studies of rectangular grooves,
we adopt a model with a long-range wall-fluid potential and fix the internal angle to $\phi=\pi/2$. We seek for the dependence of the local height of the
adsorbed liquid film above the edge $\ell_E$ on the chemical potential offset from saturation $\delta\mu\equiv\mu_{\rm s}(T)-\mu$ when the bulk coexistence is
approached from below, $\delta\mu\to 0^+$. To this end, we consider two substrate models, as schematically pictured in Fig.~\ref{sketch}. Using Model 1, the
effective Hamiltonian theory  reveals that
 \bb
\ell_E(0)-\ell_E(\delta\mu)\sim \delta \mu^{\beta_E^{co}}+{\cal{O}}(\delta\mu)\,,\label{ell_e}
 \ee
as $\delta\mu\to0^+$ with a non-universal critical exponent $\beta_E^{co}=p/(p+1)$, where the parameter $p$ characterises a decay of the binding potential far
from the edge $W(\ell)\sim \ell^{-p}$ (for $\delta\mu=0$). In the most relevant case of (3D) non-retarded van der Waals forces $p=2$, whence $\beta_E^{co}=2/3$.
In contrast, the next-to-leading term in (\ref{ell_e}) scales linearly with $\delta\mu$, regardless of the nature of the molecular interaction. We confirm this
prediction by the numerical solution of a microscopic DFT. However, for small $\delta\mu$, the requirements on the system size become rather challenging.
Therefore, as an alternative, we also consider Model 2 with a finite wall of a square cross-section with a linear dimension $L$ and use scaling arguments to
relate the height of the interface above the edge $\ell(L)$ with the wall size:
 \bb
\ell_E-\ell_E(L)\sim L^{-1}+{\cal{O}}(L^{-\frac{3}{2}})\,,\label{L}
 \ee
where all powers now depend on the molecular model and can be expressed in terms of the critical exponents characterising wetting on a planar wall. This
prediction is also confirmed by the DFT, whose implementation for Model 2 is rather straightforward.

We conclude this section by briefly recalling some properties of complete wetting on a planar wall for 3D systems with long-range forces (see, e.g.,
Ref.~\cite{dietrich}) that are relevant for our purposes. We fix the temperature to a value between the wetting temperature $T_w$ and the bulk critical
temperature $T_c$ and consider the limit $\delta\mu\to0^+$. The mean thickness of the wetting layer $\ell_\pi(\delta\mu)$ is driven by the effective interaction
(binding potential) between the wall surface and the liquid-vapour interface:
 \bb
 W(\ell)=\delta\mu\Delta\rho\ell+B\ell^{-p}+\cdots \label{vb}
 \ee
where $B>0$ is the Hamaker constant and $\Delta\rho=\rho_l-\rho_v$ is the difference between the liquid density and the vapour density at the bulk coexistence.
The global minimum of $ W(\ell)$ is at the finite value of $\ell_\pi$ for any $\delta\mu>0$, but as $\delta\mu\to0^+$, $\ell_\pi$ continuously diverges. The
singularity of $W(\ell_\pi)$ at $\delta\mu=0$ can be characterised by the set of critical exponents, in particular \cite{lipowsky}:
 \begin{eqnarray}
 \ell_\pi\sim\delta\mu^{-\beta_s^{co}}\,,\;\;\;\beta_s^{co}&=&1/(p+1)\,,\label{ell_sing}\\
 \xi_{||}\sim\delta\mu^{-\nu_{||}^{co}}\,,\;\;\;\nu_{||}^{co}&=&(p+2)/(2p+2)\,,\label{xi_sing}\\
 f_{\rm sing}\sim\delta\mu^{2-\alpha_s^{co}}\,,\;\;\;\alpha_s^{co}&=&(p+2)/(p+1)\,, \label{fsing}
 \end{eqnarray}
where $\xi_{||}$ is the transverse correlation length, and $f_{\rm sing}$ denotes a singular part of the surface free energy. We recall that the upper critical
dimension for complete wetting is $d_u<3$ for any finite value of $p$, so that the expressions (\ref{ell_sing})--(\ref{fsing}) are also valid beyond the
mean-field approximation in our three-dimensional system \cite{lipowsky}.

The remainder of the paper is organised as follows. In section 2, we describe our DFT model. An effective Hamiltonian theory and the finite-size scaling
arguments are presented in section 3, and their predictions are compared with the DFT in section 4. The results are summarised and discussed in section 5.

\section{Density Functional Theory}

%\begin{figure}
%\includegraphics[width=0.4\textwidth]{M1.png}
%\caption{A sketch of the Model 1 in the $x$-$z$ projection. As denoted, the liquid film thickness above the edge of the substrate is $\ell_E$, while far away
%from the edge the film thickness corresponds to the one above the flat wall. The wall is assumed to be infinitely long along the $y$-coordinate.} \label{fig1}
%\end{figure}

In the classical density functional theory \cite{evans79}, the equilibrium density profile minimises the grand potential functional
 \bb
 \Omega[\rho]={\cal F}[\rho]+\int\dd\rr\rhor[V(\rr)-\mu]\,,\label{om}
 \ee
where $\mu$ is the chemical potential, and $V(\rr)$ is the external potential. The intrinsic free energy functional ${\cal F}[\rho]$ can be separated into an
exact ideal gas contribution and an excess part:
  \bb
  {\cal F}[\rho]=\beta^{-1}\int\dr\rho(\rr)\left[\ln(\rhor\Lambda^3)-1\right]+{\cal F}_{\rm ex}[\rho]\,,
  \ee
where $\Lambda$ is the thermal de Broglie wavelength and $\beta=1/k_BT$ is the inverse temperature. As is common in the modern DFT approaches, the excess part is
modelled as a sum of hard-sphere and attractive contributions where the latter is treated in a simple mean-field fashion:
  \bb
  {\cal F}_{\rm ex}[\rho]={\cal F}_{\rm hs}[\rho]+\frac{1}{2}\int\int\dd\rr\dd\rr'\rhor\rho(\rr')u_{\rm a}(|\rr-\rr'|)\,, \label{f}
  \ee
where  $u_{\rm a}(r)$ is the attractive part of the fluid-fluid interaction potential.

Minimisation of (\ref{om}) leads to an Euler-Lagrange equation
 \bb
 V(\rr)+\frac{\delta{\cal F}_{\rm hs}[\rho]}{\delta\rho(\rr)}+\int\dd\rr'\rho(\rr')u_{\rm a}(|\rr-\rr'|)=\mu\,.\label{el}
 \ee

The fluid atoms are assumed to interact with one another via the truncated (i.e., short-ranged) and non-shifted Lennard-Jones-like potential
 \bb
 u_{\rm a}(r)=\left\{\begin{array}{cc}
 0\,;&r<\sigma\,,\\
-4\varepsilon\left(\frac{\sigma}{r}\right)^6\,;& \sigma<r<r_c\,,\\
0\,;&r>r_c\,.
\end{array}\right.\label{ua}
 \ee
which is cut-off at $r_c=2.5\,\sigma$, where $\sigma$ is the hard-sphere diameter.

The hard-sphere part of the excess free energy is approximated using the FMT functional \cite{ros},
 \bb
{\cal F}_{\rm hs}[\rho]=\frac{1}{\beta}\int\dd\rr\,\Phi(\{n_\alpha\})\,,\label{fmt}
 \ee
which accurately takes into account the short-range correlations between fluid particles. From the number of various FMT versions (see, e.g., Ref. \cite{roth}),
we have adopted the original Rosenfeld theory.

The wall atoms, which are assumed to be uniformly distributed with a density of $\rho_w$,  interact with the fluid particles via the Lennard-Jones--like
potential
 \bb
 \phi(r)=-\frac{4\varepsilon_w\sigma^6}{(r^2+\sigma^2)^3}\,, \label{wall}
 \ee
where $r$ is the distance between the fluid and the wall atoms.

In the following, two substrate models (walls) are considered. Within Model 1, the external potential $V(\rr)$ is induced by two semi-infinite planes that meet
at an angle $\phi=\pi/2$ as sketched in Fig.~\ref{sketch} (left). The wall is assumed to be impenetrable for the fluid particles, so that
 \bb
 V_1(x,z)=\left\{\begin{array}{ll} \infty\,; & x>0 \wedge z<0\,,\\
 \tilde{V_1}(x,z)\,; &{\rm otherwise\,,}\end{array}\right.
 \ee
which defines the attractive part of the wall potential $\tilde{V_1}(x,z)$. Assuming the translation invariance of the system along the edge, $\tilde{V_1}$ is
given by integrating over the entire depth of the wall:
 \begin{eqnarray}
\tilde{V_1}(x,z)&=&\rho_w\int_0^\infty dx'\int_{-\infty}^\infty dy'\int_{-\infty}^0 dz'\nonumber\\
&&\phi\left(\sqrt{(x-x')^2+y'^2+(z-z')^2}\right)\,,
%&=&-\frac{3}{2}\pi\varepsilon_w\sigma^6\rho_w\int_0^\infty dx'\int_{-\infty}^0 dz'\frac{1}{[(x-x')^2+(z-z')^2+\sigma^2]^{\frac{5}{2}}}\\
%&=&-\frac{1}{2}\pi\varepsilon_w\sigma^6\rho_w\int_{-\infty}^0
%dz'\left[2x^3+\sigma^2(3x+2\sqrt{\sigma^2+x^2+(z-z')^2})+2x^2\sqrt{\sigma^2+x^2+(z-z')^2}+3x(z-z')^2\right.\\
%&&\left.+2\sqrt{\sigma^2+x^2+(z-z')^2}(z-z')^2\right]\left\{[\sigma^2+(z-z')^2]^2[\sigma^2+x^2+(z-z')]^{\frac{3}{2}}\right\}^{-1}\\
%&=&-\frac{1}{4}\pi\varepsilon_w\sigma^3\rho_w\left\{\pi+\frac{2\sigma\left[\sigma^32(x-z)+xz\left(z-x-\frac{2\sigma^2+x^2+z^2}{\sqrt{\sigma^2+x^2+z^2}}\right)\right]}{(\sigma^2+x^2)(\sigma^2+z^2)}
%  +2\arctan\left(\frac{x}{\sigma}\right)-2\arctan\left(\frac{z}{\sigma}\right)\right.\\
%  &&\left.-2\arctan\left(\frac{xz}{\sigma\sqrt{\sigma^2+x^2+z^2}}\right)\right\}\\\\
%V(0,z>0)&=&-\frac{1}{4}\varepsilon_w\rho_w\sigma^6\frac{\pi^2\sigma^2-2\pi\arctan\left(\frac{z}{\sigma}\right)\sigma^2+\pi^2z^2-2\pi\sigma z-2\pi\arctan\left(\frac{z}{\sigma}\right)z^2}{\sigma^5+\sigma^3z^2}\\
%V(0,z\gg\sigma)&\approx&-\frac{1}{3}\pi\varepsilon_w\rho_w\sigma^6\frac{1}{z^3}+{\cal{O}}(z^{-5})\\
%V(0,0)&=&-\frac{1}{4}\pi^2\varepsilon_w\rho_w\sigma^3
 \end{eqnarray}
which upon substitution from (\ref{wall}), results in
 \begin{widetext}
 \begin{eqnarray}
\tilde{V_1}(x,z)&=&-\frac{1}{4}\pi\varepsilon_w\sigma^3\rho_w \left\{\pi+2\arctan\left(\frac{x}{\sigma_w}\right)\label{V}
+\frac{2\sigma\left[\sigma^2(x-z)+xz\left(z-x-\frac{2\sigma^2+x^2+z^2}{\sqrt{\sigma^2+x^2+z^2}}\right)\right]}{(\sigma^2+x^2)(\sigma^2+z^2)}
-2\arctan\left(\frac{z}{\sigma}\right)\right.\nonumber\\
&&\left. -2\arctan\left(\frac{xz}{\sigma\sqrt{\sigma^2+x^2+z^2}}\right)\right\}\,.\label{V1}
% \tilde{V}(0,z>0)&=&-\frac{1}{4}\varepsilon_w\rho_w\sigma^6\frac{\pi^2\sigma^2-2\pi\arctan\left(\frac{z}{\sigma}\right)\sigma^2+\pi^2z^2-2\pi\sigma z-2\pi\arctan\left(\frac{z}{\sigma}\right)z^2}{\sigma^5+\sigma^3z^2}\\
% \tilde{V}(0,z\gg\sigma)&\approx&-\frac{1}{3}\pi\varepsilon_w\rho_w\sigma^6\frac{1}{z^3}+{\cal{O}}(z^{-5})\\
% \tilde{V}(0,0)&=&-\frac{1}{4}\pi^2\varepsilon_w\rho_w\sigma^3
 \end{eqnarray}
 \end{widetext}

The expression (\ref{V1}) can be compared with the potential of the planar wall based on the same molecular interaction:
 \begin{widetext}
 \begin{eqnarray}
V_{\pi}(z>0)    %&=&\rho_w\int_0^\infty dx'\int_{-\infty}^\infty dy'\int_{-\infty}^0 dz'\phi_w\left(\sqrt{(x-x')^2+y'^2+(z-z')^2}\right)\nonumber\\
=-\frac{1}{2}\pi\varepsilon_w\rho_w\sigma^3\frac{\pi(\sigma^2+z^2)-2\sigma z-2\arctan\left(\frac{z}{\sigma}\right)(\sigma^2+z^2)}{\sigma^2+z^2}\,,\label{flat}
 \end{eqnarray}
 \end{widetext}
 which has an expected $z^{-3}$ asymptotic behaviour:
 \bb
 V_{\pi}(z\gg\sigma)\approx-\frac{2}{3}\pi\varepsilon_w\rho_w\sigma^6z^{-3}+{\cal{O}}(z^{-5})\,.\label{as_flat}
 \ee

  \begin{figure}
\includegraphics[width=0.4\textwidth]{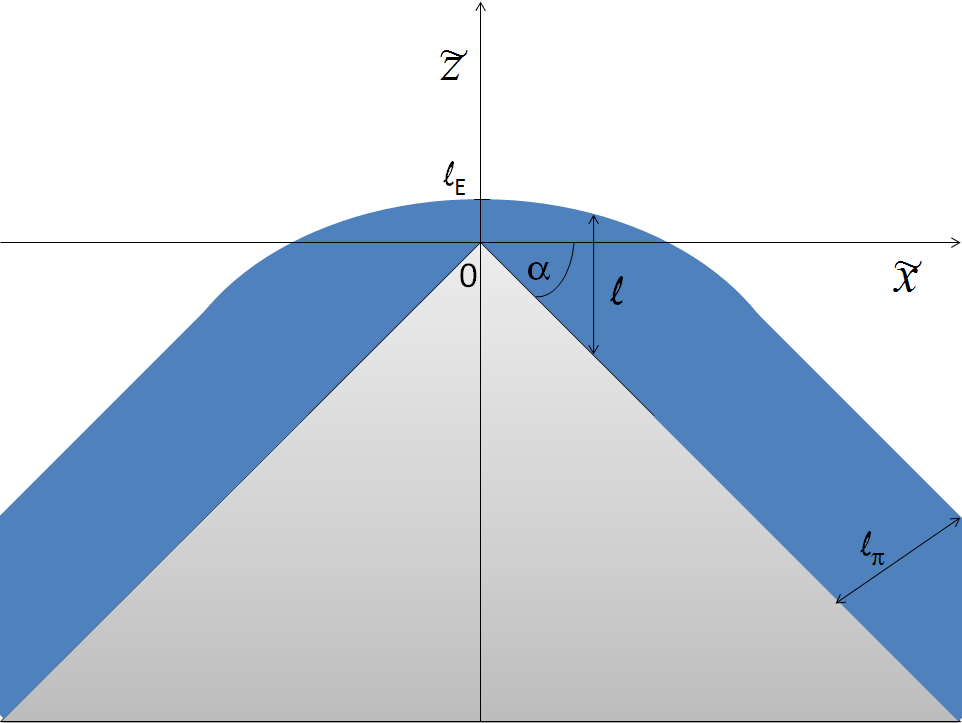}
\caption{A sketch of the substrate model that was used by the interface Hamiltonian theory. The geometry of the substrate is identical to that of Model 1, so
that the tilt angle $\alpha=\pi/4$. However, the coordinate system is now different, as depicted.} \label{EH}
\end{figure}

Within Model 2, the substrate remains assumed to be infinite along the $y$ axis, but the two other dimensions are a finite value $L$ as sketched in
Fig.~\ref{sketch} (right). In this case, the substrate potential is
 \bb
 V_2(x,z)=\left\{\begin{array}{ll} \infty\,; & |x|<L/2 \wedge |z|<L/2\,,\\
 \tilde{V_2}(x,z)\,; &{\rm otherwise\,,}\end{array}\right.
 \ee
 with
 \begin{eqnarray}
\tilde{V_2}(x,z)&=&\rho_w\int_{-L/2}^{L/2} dx'\int_{-\infty}^\infty dy' \int_{-L/2}^{L/2} dz'\\
&&\phi\left(\sqrt{(x-x')^2+y'^2+(z-z')^2}\right)\,,\nonumber
 \end{eqnarray}
leading to
 \begin{widetext}
 \begin{eqnarray}
\tilde{V_2}(x,z)=-\frac{\pi}{2}\varepsilon_w\sigma^6\rho_w\int_{-L/2-x}^{L/2-x}
dx'\left[\Psi\left(x',\frac{L}{2}-z\right)-\Psi\left(x',-\frac{L}{2}-z\right)\right]\,,
 \end{eqnarray}
  \end{widetext}
where $\Psi(x,z)=\frac{z(3x^2+2z^2+3\sigma^2)}{(x^2+\sigma^2)^2(x^2+z^2+\sigma^2)^{\frac{3}{2}}}$, which can be solved analytically.

 \begin{figure*}
\includegraphics[width=0.32\textwidth]{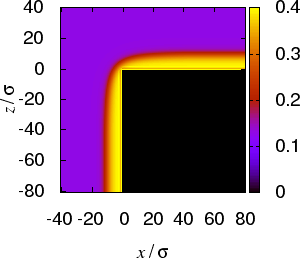} \includegraphics[width=0.32\textwidth]{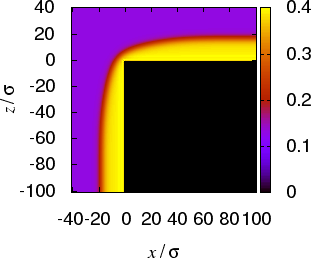} \includegraphics[width=0.32\textwidth]{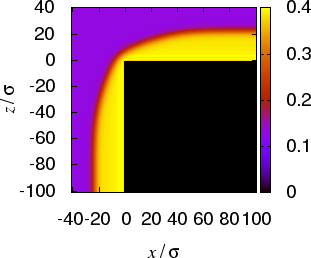}
\caption{Density profiles $\rho(x, z)$ for a fluid at the interface between a semi-infinite rectangular substrate and a bulk vapour of temperature
$k_BT/\varepsilon=1.35$ and undersaturation $(\rho_v-\rho_b)\sigma^3$ (from left to right): a) $10^{-3}$, b) $10^{-4}$ and c) $2\cdot10^{-5}$.}\label{profs_M1}
\end{figure*}

\begin{figure}
\includegraphics[width=0.5\textwidth]{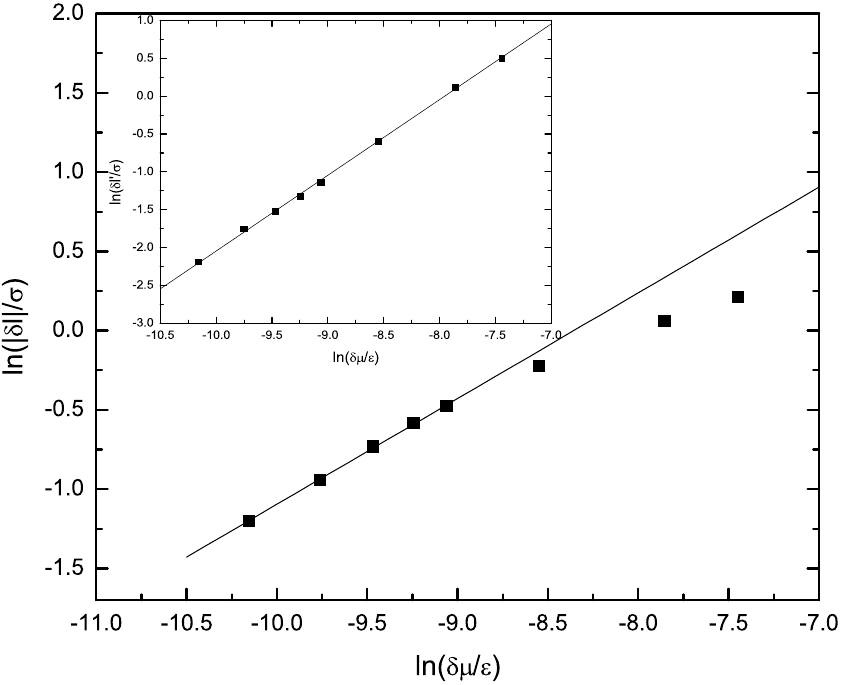}
\caption{A log-log plot of the dependence of the film thickness above the edge relative to the saturation value, $\delta\ell=\ell_E-\ell_E^0$, on the chemical
potential offset from the coexistence $\delta\mu$ for $k_BT/\varepsilon=1.35$. The symbols represent the DFT results as obtained by Model 1. The fitting line has
a gradient of $3/2$, suggesting $\delta\ell\sim\delta\mu^\frac{2}{3}$, which is consistent with Eq.~(\ref{ell_infp2}). In the inset, a log-log plot of the
dependence of $\delta\ell'=\delta\ell+C_1\delta\mu^\frac{2}{3}$ on $\delta\mu$ is shown. The fitting line has a gradient of $1$, which supports the linear form
of the second-order correction as predicted by the effective Hamiltonian theory, cf. Eq.~(\ref{ell_infp2}).} \label{dmu-ell}
\end{figure}

Using the external potentials $V_i(x,z)\,,i=1,2$, the Euler-Lagrange equations (\ref{el}) are numerically solved for the equilibrium profile $\rho(x, z)$ on a 2D
Cartesian grid with a spacing of $0.05\,\sigma$, and the corresponding integrals are performed using a Gaussian quadrature as described in Ref.~ \cite{mal}. To
model the coupling of the system with the bulk reservoir, we impose the following boundary conditions: For Model 1, we set $\rho(L_c,z>0)=\rho_\pi(z)$ and
$\rho(x<0,-L_c)=\rho_\pi(-x)$, where $L_c$ is a cut-off of the wall, and $\rho_{\pi}(z)$ is the equilibrium density profile on a corresponding planar wall. For
each bulk density (chemical potential), the grand potential minimisation is performed for different values of $L_c$ to check any possible finite-size effect on
the density distribution near the edge.
% chosen such that the height of the interface as given by the
%equilibrium (1D) density profile $\rho_{\pi}(\cdot)$ for a flat wall does not appreciably change over a reasonably large distance above the substrate.
For Model 2, we simply fix the density along the boundary of the system to the value of the vapour bulk density $\rho_b$.

\section{Interface Hamiltonian Theory and Finite Size Scaling}

From a more phenomenological perspective, the adsorption near an edge can also be studied using the interfacial Hamiltonian model \cite{wood, parry_apex}:
 \bb
 H[\ell]=\int\dd \tilde{x}\left[\frac{\gamma}{2}\left(\frac{\dd f(\tilde{x})}{\dd \tilde{x}}\right)^2+W(\ell(\tilde{x}))\right]\,.\label{H}
 \ee
The Hamiltonian is now expressed in a new Cartesian coordinate system $\{\tilde{x}, \tilde{y}, \tilde{z}\}$, which is related to the original system $\{x, y,
z\}$ by a rotation about the $y$ axis through a tilt angle $\alpha=(\pi-\phi)/2$ (see Fig~\ref{EH}); thus, the height of the wall is
$\tilde{z}_w=-\tan(\alpha)|\tilde{x}|$. Bearing in mind that for a rectangular wedge $\alpha=\pi/4$, the following analysis leaves the tilt angle unspecified.
The function $f(\tilde{x})=\ell(\tilde{x})-\tan(\alpha)|\tilde{x}|$ denotes the local height of the liquid-gas interface relative to the horizontal, and
$\ell(\tilde{x})$ is the local film thickness measured vertically.
%Notably, the shallow edge approximation $\tan\alpha\approx\alpha$ has been used.
The first term in (\ref{H}) penalises the increase of the liquid-vapour surface because of its non-planar shape, where $\gamma$ is the corresponding surface
tension, while $W(\ell)$ is the planar binding potential describing the interaction of the interface and the wall. Since the translation invariance of $\ell$
along the $y$ axis is assumed, $H[l]$ denotes the Hamiltonian of the system per unit length. We notice that $W(\ell)$ can be obtained from the DFT as a
coarse-grained excess (over bulk) grand potential (\ref{om}) using a sharp-kink approximation to the density profile \cite{dietrich}. In a mean-field
approximation, the Hamiltonian (\ref{H}) is simply minimised to yield the Euler-Lagrange equation
 \bb
 \gamma\ddot{\ell}(\tilde{x})=\frac{\partial W(\ell(\tilde{x}))}{\partial\ell}\,,
 \ee
subject to the boundary conditions $\dot{\ell}(0^+)=\tan{\alpha}$ and $\lim_{\tilde{x}\to\infty}\ell(\tilde{x})=\ell_\pi\sec\alpha$, where $\ell_\pi$ is the
equilibrium film thickness on a planar wall, and $\dot{\ell}\equiv \frac{\dd \ell(\tilde{x})}{\dd \tilde{x}}$ (note that $\ddot\ell=\ddot f$). The Euler-Lagrange
equation has a first integral, which provides an implicit equation for the height of the interface above the edge $\ell_E$:
 \bb
  \frac{\gamma\tan^2\alpha}{2}=W(\ell_E)-W(\ell_\pi\sec\alpha)\,,\label{EL}
 \ee
 which can be solved solely from knowledge of the wetting properties of the corresponding planar wall ($\alpha=0$).

At the bulk coexistence, $\ell_\pi\to\infty$ for $T>T_w$, thus the last term in Eq.~(\ref{EL}) vanishes. Then the height of the interface above the edge acquires
a simple form (cf. Ref.~\cite{parry_apex}):
 \bb
 \ell_E^0\equiv\ell_E(\delta\mu=0)=\sqrt{\frac{2B}{\gamma\tan^2\alpha}}\,,\label{ell}
 \ee
where $B$ is the Hamaker constant defined by (\ref{vb}). Because the fluid-fluid interaction is short-ranged, the Hamaker constant can be obtained from
(\ref{as_flat}):
 \bb
 W(\ell)=-\Delta\rho\int_\ell^\infty \dd z \tilde{V_\pi}(z)=\frac{B}{\ell^2}+{\cal{O}}(\ell^{-4})\,,\label{Vb}
 \ee
with $B=\frac{\pi}{3}\varepsilon_w\rho_w\sigma_w^6\Delta\rho$.

 \begin{figure*}
\includegraphics[width=0.32\textwidth]{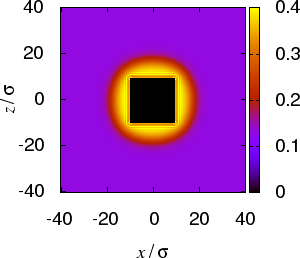} \includegraphics[width=0.32\textwidth]{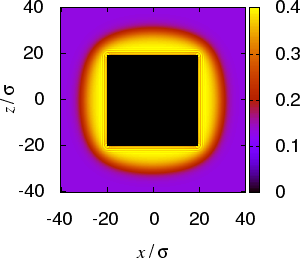} \includegraphics[width=0.32\textwidth]{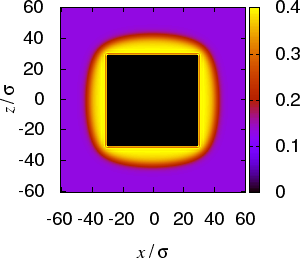}
\caption{Examples of density profiles $\rho(x, z)$ of a fluid at the interface between a bulk saturated vapour at temperature $k_BT=1.35\varepsilon$ and a
rectangular substrate of the size (from left to right): a) $L=20\sigma$, b) $L=40\sigma$ and c) $L=60\sigma$.}\label{dens_prof_L}
\end{figure*}

\begin{figure}
\includegraphics[width=0.5\textwidth]{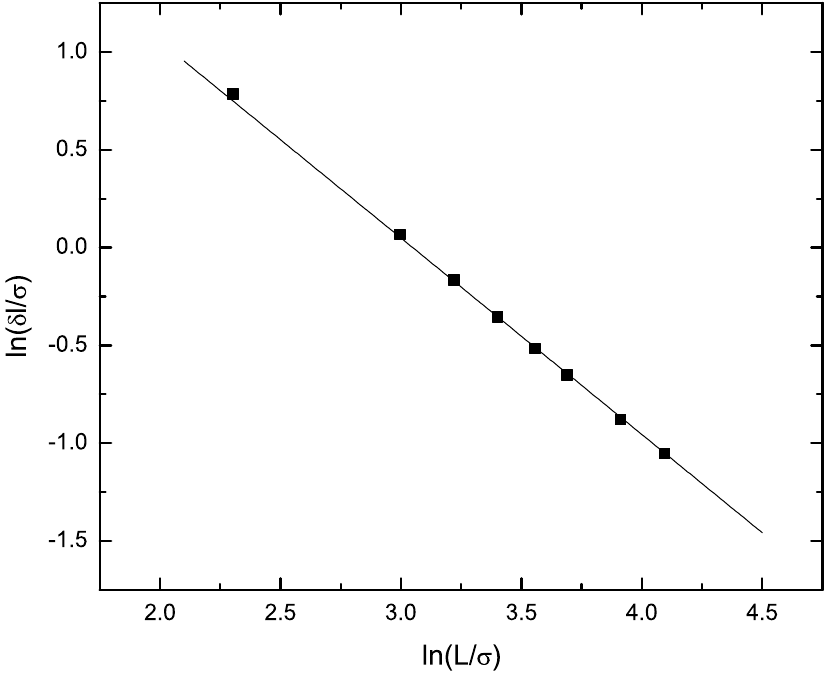}
\caption{A log-log plot of the dependence of the meniscus height above the edge on the linear dimension of the wall of size $L$. The symbols represent the
numerical DFT data corresponding to systems with saturated bulk vapour at $k_BT=1.35\varepsilon$. The gradient of the straight line fit is $-1.002$ compared to a
predicted gradient of $-1$, see Eq.~(\ref{LEseries}). } \label{L-ell}
\end{figure}

We are now concerned with the limit $\delta\mu\to0^+$. Substituting $\ell_E=\ell_E^0+\delta\ell$ into Eq.~(\ref{EL}) and introducing the abbreviation
$A=\gamma\tan^2(\alpha)/2$, one obtains
% and using the abbreviations
%$A=\gamma\tan^2(\alpha)/2$ and $\tilde{\ell}_\pi=\ell_\pi\sec\alpha$, we obtain to first order in $\delta\ell$:
 \bb
 A=\delta\mu\Delta\rho\ell_E^0+\frac{B}{(\ell_E^0)^2}-\frac{2B}{(\ell_E^0)^3}\delta\ell-W(\ell_\pi\sec\alpha)+H.O.T.
 \ee
 Using (\ref{fsing}) and (\ref{ell}), it follows that
 \bb
 \delta\ell\sim -C_1\delta\mu^{2-\alpha_s^{co}}+C_2\delta\mu\,,\label{ell_inf}
 \ee
as $\delta\mu\to 0$ and $C_1, C_2>0$. Finally, upon substituting $\alpha_s^{co}$ from Eq.~(\ref{fsing}), the exponent $\beta_E^{co}$
 defined in Eq.~(\ref{ell_e}) becomes $\beta_E^{co}=p/(p+1)$.
  More specifically, for van der Waals forces ($p=2$):
  \bb
 \delta\ell= -C_1\delta\mu^\frac{2}{3}+C_2\delta\mu+{\cal{O}}(\delta\mu^\frac{4}{3})\,.\label{ell_infp2}
 \ee

%\bb
% \delta\ell= -c_1(\delta\mu\Delta\rho)^\frac{2}{3}+c_2\delta\mu\Delta\rho+{\cal{O}}(\delta\mu^\frac{4}{3})\,.\label{ell_infp2}
% \ee
% where $c_1=3B^\frac{5}{6}/2^{\frac{5}{3}}A^{\frac{3}{2}}$ and $c_2=B/2A^2$.

In terms of Model 2, the asymptotic result (28) must be modified due to the finiteness of the linear dimension of the wall $L$ competing with the correlation
length $\xi_\parallel$. Therefore, recalling the finite-size scaling arguments (see, e.g., [32]), the result of equation (28) valid for $L\to\infty$ becomes
rescaled with a scaling function $F$:
 \begin{eqnarray}
 \delta\ell_E(L)
 &\propto&\delta\mu^{2-\alpha_s^{co}}F\left(\frac{L}{\xi_{||}}\right)+{\cal O}\left(\delta\mu F\left(\frac{L}{\xi_{||}}\right)\right)\\
 &\propto&\delta\mu^{2-\alpha_s^{co}}F(L\delta\mu^{\nu_{||}^{co}})+{\cal O}\left(\delta\mu F(L\delta\mu^{\nu_{||}^{co}})\right)\,,\nonumber
 \end{eqnarray}
 which must remain finite as $\delta\mu\to0$. Therefore,
 \begin{eqnarray}
  \delta\ell_E(L)
 &\propto&
 L^{\frac{\alpha_s^{co}-2}{\nu_{||}^{co}}}+{\cal O}\left(L^{-\frac{1}{\nu_{||}^{co}}}\right)\nonumber\\
 &\propto&L^{-1}+{\cal{O}}(L^{-\frac{3}{2}})\label{LEseries}
 \end{eqnarray}
where the values $\alpha_s^{co}=4/3$ and $\nu_{||}^{\rm co}=2/3$ were substituted in the final expression.

%In the same way, we can also determine the scaling of $\ell_M$ (see Fig.~\ref{sketch}):
% \begin{eqnarray}
% \ell_M&\sim& \ell_\pi \tilde{F}\left(\frac{\ell_\pi}{\xi_{||}}\right)
% \sim L^\frac{\beta_s^{co}}{\nu_{||}^{co}}\sim L^\frac{2}{p+2}\,, \label{LMseries}
% \end{eqnarray}
%where we have used $\ell_\pi=\ell_M(L\to\infty)\sim\delta\mu^{-\beta_s^{co}}$. For our model with $p=2$ we thus obtain $\ell_M\sim L^{1/2}$.

\section{Numerical Results}

We now examine the functional forms of (\ref{ell_infp2}) and  (\ref{LEseries})  by a comparison with the numerical solution of the microscopic DFT, as described
in section 2. We adopt $\sigma$ and $\varepsilon$ as the length and energy units, respectively, and we fix the strength of the wall potential to
$\varepsilon_w=0.4\varepsilon$, for which the wetting temperature is $k_BT_w/\varepsilon=1.25$, which is sufficiently below the bulk critical temperature
$k_BT_c/\varepsilon=1.41$. We begin with the case of a semi-infinite wall as described by Model 1. For a given value of $\delta\mu$, we first determine the
equilibrium density profile $\rho_{\pi}(z)$ for a corresponding system with a planar wall, which constitutes a boundary condition for the system with a single
edge. For the sake of numerical consistency, the profile $\rho_{\pi}(z)$, albeit varying only in one dimension, is determined on the same two-dimensional grid as
used for the edge. This also provides a good test of our numerics, since the difference between the planar density profile that is constructed from a 2D
calculation proved not to appreciably differ from that obtained from a standard 1D treatment. Moreover, the numerical accuracy of the full 2D DFT code, described
in details in Ref.~\cite{mal}, was verified by comparison of the DFT results with the exact pressure sum-rule \cite{mal_parry_wedge}. Then we set the boundary
conditions such that $\rho(L_c,z>0)=\rho_\pi(z)$ and $\rho(x<0,-L_c)=\rho_\pi(-x)$, where the value of the wall cut-off $L_c$ ranges from $L_c=40\sigma$ to
$L_c=100\sigma$ to verify that the system size does not affect $\ell_E$.

The representative samples of the equilibrium density profiles are shown in Fig.~\ref{profs_M1}. The height of the fluid interface above the edge is defined as
follows:
 \bb
\ell_E=\frac{\sqrt{2}}{\Delta\rho}\int_{-\infty}^0\dd x\left(\rho(x,-x)-\rho_b\right)\,, \label{ell_E}
 \ee
where $\rho_b$ is the density of the gas reservoir. Eq.~(\ref{ell_E}) allows us to compare the DFT results with the prediction based on the interface Hamiltonian
theory as given by (\ref{ell_infp2}). The comparison that is displayed in Fig.~\ref{dmu-ell} reveals a consistency between the two approaches and verifies the
values of the exponents of the two first terms in Eq.~(\ref{ell_infp2}).
%especially for small values of $\delta\mu$, where the terms beyond the order linear in $\delta\mu$, that we neglect in the comparison, become unimportant.

Next, we consider Model 2 and examine the validity of the expansion (\ref{LEseries}). In the DFT, the density at the boundary of the system is fixed to the value
of the bulk density of the saturated vapour, $\rho_b=\rho_v$, and the linear dimension of the box size is chosen from a range between $80\sigma$ and $120\sigma$.
Varying the wall size $L$, we find the equilibrium state of each system as shown in Fig.~\ref{dens_prof_L}. In Fig.~\ref{L-ell}, we display a log-log plot of the
height of the interface above the edge $\ell_E$ versus the wall size $L$. The values of $\ell_E$ are again determined using formula (\ref{ell_E}), where the
upper limit is $-L_c/2$. The fitted line shows a good agreement between the DFT and the analytic expression (\ref{LEseries}), and for the region of $L>20\sigma$,
the first-order term in Eq.~(\ref{LEseries}) appears to dominate. The consistency between the gradient of the fitted line and the predicted value $-1$ is within
an error of $0.2\%$.

\section{Conclusion}
In this work, we used an interfacial Hamiltonian theory and a fundamental-measure DFT to study the fluid adsorption near a rectangular edge of a substrate
interacting with the fluid via van der Waals forces. When the two-phase bulk coexistence is approached from below at a fixed temperature, i.e., the deviation of
the chemical potential from the coexistence $\delta\mu=\mu_{\rm s}-\mu\to0^+$, macroscopically thick films are formed at the wall far away from the edge. Because
these asymptotic interfaces must eventually merge to form a meniscus, the local height of the interface above the edge $\ell_E$ remains finite and indeed rather
small even at the bulk coexistence. In this paper, we have shown that for an infinitely long substrate, $\ell_E(\delta\mu)$ approaches the coexistence value
according to $\delta\ell=\ell_E(0)-\ell_E(\delta\mu)\sim \delta \mu^{\beta_E^{co}}$ as $\delta\mu\to0^+$. The exponent depends on the range of the molecular
interaction, such that $\beta_E^{co}=p/(p+1)$, where $p$ defines the asymptotic decay of the binding potential $W(\ell)\sim\ell^{-p}$.  The second-order
correction to $\delta\ell$ is linear in $\delta\mu$ regardless of the molecular interaction. Both findings were verified by the DFT numerical calculations. We
also showed that if the substrate is of finite size $L$, the previous result corresponds to the scaling of $\ell_E$ as $\ell_E-\ell_E(L)\propto
L^{-1}+{\cal{O}}(L^{-\frac{3}{2}})$, as also confirmed by the DFT. We conclude with two remarks about the generality of these findings. First, throughout this
study, the substrate geometry was maintained fixed such that the substrate edge was rectangular. This geometry was selected because the model of a right-angle
edge appears important considering its connection with other fundamental substrate models as discussed in the introduction. Technically, the  external potential
for the rectangular geometry remains rather simple, which facilitates the numerics in the DFT, and the application of the finite-size arguments is
straightforward. Nevertheless, we believe that the result given by Eq.~(\ref{ell_e}) is valid for an arbitrary internal angle, since the value of $\phi$ was not
assumed in the derivation of (\ref{ell_e}). Second, because the edge geometry does not induce any new divergence compared to a planar wall, the upper critical
dimension $d_u$ corresponding to complete wetting must be identical for the two substrates. Since $d_u<3$ for a finite $p$ for a planar wall \cite{lipowsky}, our
mean-field results remain unaffected by the capillary-wave fluctuations.

\begin{acknowledgments}
 \noindent I am grateful to Andrew Parry for useful discussions. The financial support from the Czech Science Foundation, project 13-09914S, is acknowledged.
\end{acknowledgments}

\end{document}